\documentclass[12pt]{article}

\usepackage{amsmath, amssymb, amsfonts}
\usepackage{geometry}
\usepackage{graphicx}
\usepackage[colorlinks=true, allcolors=blue]{hyperref}
\usepackage{cite}
\usepackage{verbatim}
\usepackage[most]{tcolorbox}
\usepackage{tikz}
\usetikzlibrary{arrows.meta, decorations.pathmorphing}

\newcommand{\bra}[1]{\left\langle #1 \right|}
\newcommand{\ket}[1]{\left| #1 \right\rangle}
\newcommand{\braket}[2]{\left\langle #1 \middle| #2 \right\rangle}
\newcommand{\matrixel}[3]{\left\langle #1 \middle| #2 \middle| #3 \right\rangle}
\newcommand{\expvalInline}[1]{\left\langle #1 \right\rangle}
\newcommand{\expval}[1]{\left\langle #1 \right\rangle}
\newcommand{\proj}[1]{\ket{#1}\bra{#1}}
\newcommand{\Tr}{\operatorname{Tr}}
\title{\textbf{When the Weak Becomes Strong: \\ Effective Observables via Time-Symmetric Quantum Selection}}
\author{Mirco A. Mannucci \thanks{Email: mirco@holomathics.com} \\
\textit{\small HoloMathics, LLC}
}
\date{\today}

\begin{document}

\maketitle
\vspace{-1em}
\tableofcontents
\vspace{1em}

\begin{abstract}
The weak value of a quantum observable, conditioned on both preselected and post-selected states, is a cornerstone of time-symmetric quantum mechanics. We investigate the physical meaning of composing two weak values in sequence: a "forward" measurement from a preselected state $\ket{\psi}$ to a post-selected state $\ket{\phi}$, followed by a "reverse" measurement. We show that the product of these two weak values is the normalized expectation value of a strong, state-conditioned observable $B = A P_\psi A$, where $P_\psi = \ket{\psi}\bra{\psi}$ is the projection operator onto the state $\ket{\psi}$. We analyze the structure of this operator when $\ket{\psi}$ is an eigenstate of $A$ versus a superposition, revealing how it encodes interference information. This formalism provides a practical tool for quantum information science and can be generalized by replacing the pure-state projector $P_\psi$ with a generic density matrix $\rho$, connecting it to the broader framework of generalized quantum measurements. We show how these tools can be used for state-specific error witnessing in quantum computers and how the phase of the weak value can be recovered in the pure-state case entirely through strong measurements.
\vspace{1em}

\end{abstract}

\section{Introduction}

The quantum measurement process, typically described as an irreversible projection onto an eigenstate, reveals a more nuanced structure when considering systems conditioned on future information. The concept of the weak value, introduced by Aharonov, Albert, and Vaidman \cite{Aharonov1988}, formalizes this by defining an observable's value for a sub-ensemble of particles that are successfully postselected in a final state, which may differ from the initial state. These values can lie far outside the observable's eigenvalue spectrum, can be complex, and have been experimentally verified in numerous contexts \cite{Ritchie1991, Dressel2014}.

Weak values are theoretically grounded in the \textbf{Aharonov-Bergmann-Lebowitz (ABL)} formula \cite{Aharonov1964}, which gives the probability of an intermediate measurement outcome given both a preselected state $\ket{\psi}$ and a post-selected state $\ket{\phi}$. This time-symmetric perspective, central to the Two-State Vector Formalism (TSVF) \cite{Aharonov2008, Aharonov2010}, suggests that a complete description of a quantum system between two measurements requires information from both the past and the future. The TSVF has been shown to be a powerful framework for analyzing quantum phenomena, leading to new insights and even stimulating discoveries in other fields like mathematics \cite{Aharonov2009, Tollaksen2007}.

While the forward weak value is widely studied, the physical significance of combining it with its time-reversed counterpart is less explored. This paper addresses a fundamental question: \textit{What observable quantity is represented by the product of a forward and a reverse weak value?}

We show that this product corresponds to the expectation value of a well-defined, state-dependent operator $B = A P_\psi A$, where $P_\psi =\ket{\psi}\bra{\psi}$. This operator, which we term an "effective", or perhaps more accurately a "state-conditioned" observable, is quadratic in the original observable $A$ and depends explicitly on the intermediate state $\ket{\psi}$. We analyze the structure of $B$ in detail and generalize it to the case where the conditioning system is in a mixed state $\rho$, yielding the operator $B_\rho = A \rho A$. We explore the properties of these operators and propose their use as tools for context-aware diagnostics in quantum computing.

This paper is structured as follows. In Section \ref{sec:formalism}, we review the formalism of weak values. In Section \ref{sec:bidirectional}, we derive our main result, introduce the effective operator $B$, and analyze its structure. In Section \ref{sec:phase}, we demonstrate how to recover the weak value's phase entirely via strong measurements. Section \ref{sec:applications} discusses applications, with a focus on quantum computing, and future directions, including the generalization to mixed states. Finally, we provide a Qiskit simulation in the Appendix to verify our claims for both pure and mixed states and for further hands-on explorations.

\section{ The Weak Value}
\label{sec:formalism}

Let $A$ be a Hermitian operator on a Hilbert space $\mathcal{H}$. Consider an ensemble of quantum systems prepared in an initial (preselected) state $\ket{\psi}$. A final (post-selected) measurement is performed, retaining only those systems found in the state $\ket{\phi}$.

The weak value of the observable $A$ for this sub-ensemble is defined as \cite{Aharonov1988}:
\begin{equation}
    A_w := \frac{\matrixel{\phi}{A}{\psi}}{\braket{\phi}{\psi}}.
    \label{eq:weak_value_def}
\end{equation}
This definition requires that the pre-selected and post-selected states be non-orthogonal, i.e., $\braket{\phi}{\psi} \neq 0$.\footnote{The formalism diverges when the states are orthogonal. This singularity is the mathematical foundation of weak value amplification, where choosing a nearly orthogonal post-selection can lead to anomalously large weak values, provided the numerator $\matrixel{\phi}{A}{\psi}$ does not also vanish.} In general, $A_w$ is a complex number. Its real part is interpreted as the average shift in a measurement device's pointer, and its imaginary part is related to the shift in the conjugate momentum of the pointer \cite{Jozsa2007}. The experimental procedure for extracting such values is known as a \textbf{weak measurement}, which can be made robust for finite samples \cite{Tollaksen2007_robust}.

\section{Bidirectional Weak Values and the Effective Observable}
\label{sec:bidirectional}
The focal point of this article is deceptively simple: imagine you have fixed a pre-selected state and a post-selected one, now do a double trip: for the first to the next and back, meanwhile using the particular lens $A$ (see Figure below). What do you get?
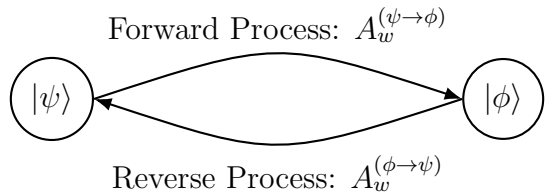
\begin{figure}[h!]
\centering
\begin{tikzpicture}[
    state/.style={circle, draw, thick, minimum size=1cm},
    arrow/.style={-Latex, thick},
    mirror/.style={draw, thick, fill=gray!20, decoration={zigzag, segment length=5, amplitude=1.5}, decorate}
]
    \node[state] (psi) at (0,0) {$\ket{\psi}$};
    \node[state] (phi) at (6,0) {$\ket{\phi}$};

    \draw[arrow] (psi.east) .. controls (3, 0.8) .. (phi.west)
        node[midway, above] {Forward Process: $A_w^{(\psi \to \phi)}$}
        ;

    \draw[arrow] (phi.west) .. controls (3, -0.8) .. (psi.east)
        node[midway, below] {Reverse Process: $A_w^{(\phi \to \psi)}$}
        ;

\end{tikzpicture}
\caption{A schematic of the bidirectional weak value process. The system evolves from a preselected state $\ket{\psi}$ to a post-selected state $\ket{\phi}$ (forward process). At $\ket{\phi}$, a metaphorical "time mirror" reverses the process, describing the evolution from $\ket{\phi}$ back to $\ket{\psi}$ (reverse process). The product of these two weak values corresponds to the expectation value of the effective operator $B = A P_\psi A$.}
\label{fig:time_mirror}
\end{figure}

\subsection{The Product of Forward and Reverse Weak Values}

The "forward" weak value is given by Eq. \eqref{eq:weak_value_def}. The "reverse" weak value is $A_w^{(\phi \to \psi)} := \frac{\matrixel{\psi}{A}{\phi}}{\braket{\psi}{\phi}}$.
Since $A$ is Hermitian, their product is:
\begin{equation}
    C_{\psi \phi} := A_w^{(\psi \to \phi)} \cdot A_w^{(\phi \to \psi)} = \frac{|\matrixel{\phi}{A}{\psi}|^2}{|\braket{\phi}{\psi}|^2} = |A_w^{(\psi \to \phi)}|^2.
    \label{eq:product}
\end{equation}
The quantity $C_{\psi \phi}$ is real and non-negative. It is worth noting the deep symmetry here. While our quantity $C_{\psi \phi}$ is the product of the forward and reverse weak values, the *real part* of the weak value, often the primary quantity measured in experiments, is their arithmetic mean: $\text{Re}(A_w^{(\psi \to \phi)}) = \frac{1}{2}(A_w^{(\psi \to \phi)} + A_w^{(\phi \to \psi)})$.

\subsection{The Effective Operator \texorpdfstring{$B = A P_\psi A$}{B = A P_psi A}}

Let $P_\psi := \ket{\psi}\bra{\psi}$ be the projection operator onto the state $\ket{\psi}$. We define the effective operator $B$ as:
\begin{equation}
    B := A P_\psi A.
    \label{eq:B_def}
\end{equation}
This operator is Hermitian and positive semi-definite. Its expectation value in the post-selected state $\ket{\phi}$ is precisely the numerator of Eq. \eqref{eq:product}:
\begin{equation}
    \matrixel{\phi}{B}{\phi} = \matrixel{\phi}{A P_\psi A}{\phi} = \matrixel{\phi}{A}{\psi}\matrixel{\psi}{A}{\phi} = |\matrixel{\phi}{A}{\psi}|^2.
    \label{eq:expval_B}
\end{equation}
With this result, we can now express the weak value product $C_{\psi\phi}$ directly in terms of standard expectation values. The denominator of Eq. \eqref{eq:product} is

$|\braket{\phi}{\psi}|^2 = \braket{\phi}{\psi} \braket{\psi}{\phi} = \matrixel{\phi}{P_\psi}{\phi}$

Substituting this and Eq. \eqref{eq:expval_B} into Eq. \eqref{eq:product} gives:
\begin{equation}
    C_{\psi \phi} = \frac{\matrixel{\phi}{B}{\phi}}{\matrixel{\phi}{P_\psi}{\phi}}.
\end{equation}
This equation shows how the product of two time-symmetric weak values gives rise to a quantity measurable as a ratio of two conventional, strong expectation values.

\begin{tcolorbox}[colback=gray!5!white, colframe=gray!50!black, title=Recovering the Modulus of the Weak Value]
We note an elegant consequence: since the product of the forward and reverse weak values is given by
\[
C_{\psi\phi} = \left| A^{(\psi \to \phi)}_w \right|^2 = \frac{ \matrixel{\phi}{B}{\phi} }{ \matrixel{\phi}{P_\psi}{\phi} },
\]
it follows immediately that the \emph{modulus} of the weak value can be expressed as:
\[
\left| A^{(\psi \to \phi)}_w \right| = \sqrt{ \frac{ \matrixel{\phi}{B}{\phi} }{ \matrixel{\phi}{P_\psi}{\phi} } } = \sqrt{ \frac{ \matrixel{\phi}{AP_\psi A}{\phi} }{ \matrixel{\phi}{P_\psi}{\phi} } }.
\]
Thus, even though the weak value is complex and defined via a quotient of amplitudes, its \emph{modulus} arises entirely from strong measurements of two Hermitian operators: \( B \) and \( P_\psi \).
\end{tcolorbox}

\subsection{Analysis of the Operator B}
The structure of $B$ depends critically on the relationship between the conditioning state $\ket{\psi}$ and the observable $A$.

\textbf{Case 1: $\ket{\psi}$ is an eigenstate of A.}
Let $\ket{\psi}$ be an eigenstate of $A$ with a real eigenvalue $a$, such that $A\ket{\psi} = a\ket{\psi}$. The operator $B$ simplifies dramatically:
\begin{equation}
    B = A P_\psi A = A (\ket{\psi}\bra{\psi}) A = (A\ket{\psi})(A\ket{\psi})^\dagger = (a\ket{\psi})(a\ket{\psi})^\dagger = a^2 \ket{\psi}\bra{\psi} = a^2 P_\psi.
\end{equation}
In this case, the effective operator $B$ is simply the projector onto the eigenstate $\ket{\psi}$, scaled by the eigenvalue squared.

\textbf{Case 2: $\ket{\psi}$ is a superposition of eigenstates.}
Let $\ket{\psi} = \sum_i c_i \ket{a_i}$, where $A\ket{a_i} = a_i\ket{a_i}$. The action of $A$ on $\ket{\psi}$ produces the state $A\ket{\psi} = \sum_i c_i a_i \ket{a_i}$. The operator $B$ is the projector onto this new state:
\begin{equation}
    B = (A\ket{\psi})(A\ket{\psi})^\dagger = \left( \sum_i c_i a_i \ket{a_i} \right) \left( \sum_j c_j^* a_j \bra{a_j} \right) = \sum_{i,j} c_i c_j^* a_i a_j \ket{a_i}\bra{a_j}.
\end{equation}
This form is much richer. It contains diagonal terms ($i=j$), which represent populations, and off-diagonal terms ($i \neq j$), which are coherences in the eigenbasis of $A$. These terms encode the interference effects resulting from the action of $A$ on the superposition state $\ket{\psi}$.

\section{Phase Recovery for Pure States via Strong Measurements}
\label{sec:phase}

The product $C_{\psi \phi}$ is real and thus discards the phase of the weak value. We now show that this phase is also accessible entirely through strong measurements. We begin by defining the non-Hermitian operator $\mathcal{C} := A P_\psi$. Its expectation value in $\ket{\phi}$ is $\matrixel{\phi}{\mathcal{C}}{\phi} = \matrixel{\phi}{A P_\psi}{\phi} = \matrixel{\phi}{A}{\psi} \braket{\psi}{\phi}$. Comparing this to the definition of the weak value, we find:
\begin{equation}
    A_w^{(\psi \to \phi)} = \frac{\matrixel{\phi}{\mathcal{C}}{\phi}}{|\braket{\phi}{\psi}|^2},
\end{equation}
which implies that the phase of the weak value is given by $\text{arg}(A_w) = \text{arg}(\matrixel{\phi}{\mathcal{C}}{\phi})$.

The challenge is that $\mathcal{C}$ is not an observable. However, its expectation value can be determined by measuring the expectation values of its Hermitian and anti-Hermitian parts. Any operator can be decomposed as $\mathcal{C} = \mathcal{C}_R + i\mathcal{C}_I$, where the components are the two Hermitian operators:
\begin{align}
    \mathcal{C}_R &= \frac{1}{2}(\mathcal{C} + \mathcal{C}^\dagger) = \frac{1}{2}(A P_\psi + P_\psi A) \\
    \mathcal{C}_I &= \frac{1}{2i}(\mathcal{C} - \mathcal{C}^\dagger) = \frac{1}{2i}[A, P_\psi]
\end{align}
Since $\matrixel{\phi}{\mathcal{C}}{\phi} = \matrixel{\phi}{\mathcal{C}_R}{\phi} + i\matrixel{\phi}{\mathcal{C}_I}{\phi}$, we can experimentally determine the complex number $\matrixel{\phi}{\mathcal{C}}{\phi}$ and thus the weak value's phase.

This leads to the following experimental protocol:
\begin{enumerate}
    \item \textbf{Construct Observables (Classical Step):} Given $A$ and $\ket{\psi}$, mathematically construct the two Hermitian operators (observables) $\mathcal{C}_R$ and $\mathcal{C}_I$. For qubit systems, this involves finding their Pauli decompositions.
    \item \textbf{Measure $\matrixel{\phi}{\mathcal{C}_R}{\phi}$ (Experimental Step):} Prepare an ensemble of systems in the state $\ket{\phi}$ and perform strong measurements to determine the expectation value of $\mathcal{C}_R$. Let the result be the real number $x = \matrixel{\phi}{\mathcal{C}_R}{\phi}$.
    \item \textbf{Measure $\matrixel{\phi}{\mathcal{C}_I}{\phi}$ (Experimental Step):} Prepare another ensemble in the state $\ket{\phi}$ and perform strong measurements to determine the expectation value of $\mathcal{C}_I$. Let the result be the real number $y = \matrixel{\phi}{\mathcal{C}_I}{\phi}$.
    \item \textbf{Calculate the Phase (Classical Step):} The phase of the weak value is the argument of the complex number you have just constructed:
    \begin{equation}
        \text{Phase of } A_w = \text{arg}(x + iy) = \text{atan2}(y, x).
    \end{equation}
\end{enumerate}

\begin{figure}[h!]
\centering
\begin{tikzpicture}[
    box/.style={rectangle, draw, thick, minimum width=3.2cm, minimum height=1cm, align=center},
    arrow/.style={->, thick}
]

\node[box] (phi) at (0,0) {$\ket{\phi}$};

\node[box] (CR) at (5,1.5) {Measure $\mathcal{C}_R$ };
\node[box] (CI) at (5,-1.5) {Measure $\mathcal{C}_I$ };

\node[box] (combine) at (10,0) {$\matrixel{\phi}{\mathcal{C}}{\phi} = x + i y$ \\ phase is $\text{arg}(x + i y)$};

\draw[arrow] (phi) -- (CR);
\draw[arrow] (phi) -- (CI);
\draw[arrow] (CR) -- (combine);
\draw[arrow] (CI) -- (combine);

\end{tikzpicture}
\caption{Experimental procedure for determining the phase of the weak value $A_w$ using two strong measurements: one of the observable $\mathcal{C}_R$ and one of $\mathcal{C}_I$. Combining the results allows reconstruction of the complex quantity $\matrixel{\phi}{\mathcal{C}}{\phi}$, from which the phase is extracted using the operation $\text{arg}(x + i y)$.}
\label{fig:phase_recovery}
\end{figure}
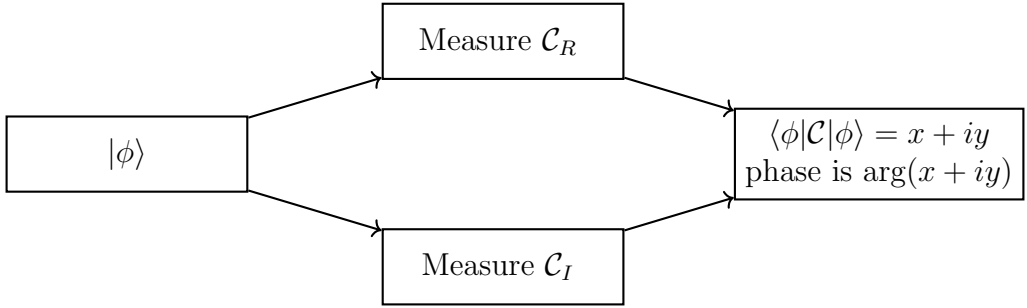

\begin{tcolorbox}[colback=gray!5!white, colframe=gray!50!black, title=Physical Interpretation of the Operators $\mathcal{C}_R$ and $\mathcal{C}_I$]
The Hermitian operators $\mathcal{C}_R = \frac{1}{2}\{A, P_\psi\}$ (the anti-commutator) and $\mathcal{C}_I = \frac{1}{2i}[A, P_\psi]$ (the commutator) encode two complementary aspects of the interaction between the observable $A$ and the preselected state $\ket{\psi}$:
\begin{itemize}
    \item \textbf{$\mathcal{C}_R$} captures the \textit{symmetric, classical-like component} of the interaction: it quantifies how much the action of $A$ reinforces or preserves the structure of the state $\ket{\psi}$.
    \item \textbf{$\mathcal{C}_I$} measures the \textit{asymmetric, quantum component}: it reflects the degree of non-commutativity, and thus the potential disturbance or complementarity between the observable and the state.
\end{itemize}
\end{tcolorbox}

This protocol demystifies the phase of the weak value, showing that it is not an exotic quantity but is encoded in the relative expectation values of two standard observables. This completes our picture: the squared modulus of the weak value corresponds to the operator $B = A P_\psi A$ (which is equal to $\mathcal{C}^\dagger\mathcal{C}$), while its phase is determined by the two Hermitian parts of $\mathcal{C}$.

\section{Applications }
\label{sec:applications}

\subsection{Application to Quantum Computing: State-Specific Error Witnessing}

Our formalism provides a tool for \textbf{state-specific error witnessing}. Consider a gate $U_A = e^{-i\theta A}$ that acts on a critical input state $\ket{\psi}$. We can diagnose errors in this specific operation using the operator $B = A P_\psi A$ as a targeted witness. The protocol is:
\begin{enumerate}
    \item \textbf{Construct the Witness:} From the gate generator $A$ and the state $\ket{\psi}$, form the operator $B = A P_\psi A$.
    \item \textbf{Calculate Ideal Value:} Theoretically calculate the expected outcome for a perfect gate, $\expvalInline{B}_{\text{ideal}} = \matrixel{U_A\psi}{B}{U_A\psi}$.
    \item \textbf{Measure Experimental Value:} On the quantum computer, prepare $\ket{\psi}$, apply the actual noisy gate $\mathcal{E}$, and measure the expectation value of the witness $B$, yielding 
    $\expvalInline{B}_{\text{real}} = \Tr\left(B\, \mathcal{E}(\proj{\psi})\right)$.
\end{enumerate}
 The deviation $\Delta = |\expval{B}_{\text{ideal}} - \expval{B}_{\text{real}}|$ provides a sensitive, context-aware measure of the gate's error as it pertains to its action on $\ket{\psi}$.

\subsection{Generalization to Mixed States}
The formalism can be extended by replacing the pure-state projector $P_\psi = \proj{\psi}$ with a general density matrix $\rho$, representing a mixed preselected state. We define a generalized effective operator:
\begin{equation}
    B_\rho := A \rho A.
\end{equation}
This operator is also Hermitian and positive semi-definite. Let the spectral decomposition of the preselection state be $\rho = \sum_k p_k \proj{\psi_k}$, where $p_k$ are the probabilities for the system to be in the pure state $\ket{\psi_k}$. The expectation value of $B_\rho$ in a final state $\ket{\phi}$ is:
\begin{align}
    \matrixel{\phi}{B_\rho}{\phi} &= \matrixel{\phi}{A \left( \sum_k p_k \proj{\psi_k} \right) A}{\phi} \nonumber \\
    &= \sum_k p_k \matrixel{\phi}{A \proj{\psi_k} A}{\phi} = \sum_k p_k |\matrixel{\phi}{A}{\psi_k}|^2.
\end{align}
This has a clear physical interpretation: it is the \textit{incoherent average of the squared transition amplitudes}, weighted by the classical probabilities of the constituent pure states in the mixture $\rho$.

\begin{tcolorbox}[colback=gray!5!white, colframe=gray!50!black, title= Interpreting $B_\rho$: Weak-Value Structure in the Mixed-State Regime]
This generalization connects our work to the broader theory of quantum operations and POVMs. An operation of the form $\rho \to A\rho A$ is a valid (though not trace-preserving) quantum operation.
\\
While the direct link to a product of weak values is obscured—as a mixed state does not have a single weak value, but an averaged one—the operator $B_\rho$ generalizes the concept of the squared modulus to scenarios involving mixed preselected states. The phase recovery, however, does not have a straightforward generalization, as it would require a coherent averaging of complex numbers.
\end{tcolorbox}

\section{Future Directions}

Further research could explore the properties of $B_\rho$ in the context of open quantum systems, where $\rho$ might be the steady state of some dissipative process. Additionally, the quadratic nature of $B$ when $A$ is a Hamiltonian suggests potential applications in quantum thermodynamics for analyzing work fluctuations.

In conclusion, the structure $A P_\psi A$ and its generalization $A \rho A$ provide a simple yet rich conceptual and practical framework for analyzing state-conditioned quantum processes.

\section*{Acknowledgments}
I would like to thank Daniele C. Struppa and Jeff Tollaksen for kindly inviting me to Yakir Aharonov's Seminar on Weak Values at GMU in the Spring 2006. It was there and then that my interest for weak values was kindled, though it took me a long time to come back to my initial thoughts.

Special thanks are due to Domenico Napoletani for providing valuable feedback on a preliminary version of this work. His suggestion to pursue an experimental setup to ground the mathematical formalism was particularly influential.

\bibliographystyle{unsrt}

\appendix
\section{Qiskit Simulations}

The following Python scripts using Qiskit are provided to numerically verify the core results of this paper.

A complete, executable Jupyter notebook containing these simulations is available at the following GitHub repository: \\
\url{https://github.com/Mircus/quantum-weak-values-geometry}

\subsection{Simulation for the Pure State Operator}

We provide a Python script using Qiskit to numerically verify our result for pure states: $\matrixel{\phi}{B}{\phi} = |\matrixel{\phi}{A}{\psi}|^2$, where $B = A P_\psi A$. The core idea is to measure the quantity on the right-hand side using a quantum circuit. This probability can be measured by preparing $\ket{\psi}$, applying $A$, and then measuring the probability of finding the system in state $\ket{\phi}$. This is accomplished by applying the inverse of the circuit that prepares $\ket{\phi}$ (i.e., $\phi^\dagger$) and measuring the probability of being in the ground state $\ket{0}$.
\begin{verbatim}
import numpy as np
from qiskit import QuantumCircuit, transpile
from qiskit_aer import AerSimulator
from qiskit.quantum_info import Statevector, Operator

# --- 1. Define Operators and States ---
A = Operator([[1, 0], [0, -1]])
psi_circuit = QuantumCircuit(1)
psi_circuit.h(0)
psi_state = Statevector(psi_circuit)
theta_phi = np.pi / 3
phi_circuit = QuantumCircuit(1)
phi_circuit.ry(theta_phi, 0)
phi_state = Statevector(phi_circuit)

# --- 2. Theoretical Calculation ---
P_psi = Operator(np.outer(psi_state.data, psi_state.data.conj()))
B = A @ P_psi @ A
exp_B_phi = phi_state.expectation_value(B).real
amp_sq = np.abs(phi_state.data.conj() @ A.data @ psi_state.data)**2
assert np.isclose(exp_B_phi, amp_sq)
print(f"Theoretical <phi|B|phi> = {exp_B_phi:.4f}")

# --- 3. Qiskit Simulation ---
phi_dagger_circuit = phi_circuit.inverse()
circuit = QuantumCircuit(1, 1)
circuit.append(psi_circuit, [0])
circuit.append(A, [0])
circuit.append(phi_dagger_circuit, [0])
circuit.measure(0, 0)

simulator = AerSimulator()
result = simulator.run(transpile(circuit, simulator), shots=10000).result()
prob_0 = result.get_counts().get('0', 0) / 10000
print(f"Simulated |<phi|A|psi>|^2 = {prob_0:.4f}")
\end{verbatim}
\subsection{Simulation for the Mixed State Generalization}
Here, we verify the identity for mixed states: $\matrixel{\phi}{B_\rho}{\phi} = \sum_k p_k |\matrixel{\phi}{A}{\psi_k}|^2$. We choose a mixed state $\rho = 0.75 \proj{0} + 0.25 \proj{1}$. The right-hand side is an average of two squared amplitudes. We can measure it by running two separate quantum experiments—one for the case $\ket{\psi_0}=\ket{0}$ and one for $\ket{\psi_1}=\ket{1}$—and then combining their results classically with the weights $p_0=0.75$ and $p_1=0.25$.

\begin{verbatim}
import numpy as np
from qiskit import QuantumCircuit, transpile
from qiskit_aer import AerSimulator
from qiskit.quantum_info import Statevector, Operator, DensityMatrix

# --- 1. Define Operators and States ---
A = Operator([[1, 0], [0, -1]])
theta_phi = np.pi / 3
phi_circuit = QuantumCircuit(1)
phi_circuit.ry(theta_phi, 0)
phi_state = Statevector(phi_circuit)
p0, p1 = 0.75, 0.25
rho = DensityMatrix([[p0, 0], [0, p1]])
psi_k = [Statevector.from_label('0'), Statevector.from_label('1')]
pk = [p0, p1]

# --- 2. Theoretical Calculation ---
B_rho = Operator(A.data @ rho.data @ A.data)
exp_B_rho_phi = phi_state.expectation_value(B_rho).real
rhs_sum = sum(pk[i] * np.abs(phi_state.data.conj() @ A.data @ psi_k[i].data)**2 for i in range(2))
assert np.isclose(exp_B_rho_phi, rhs_sum)
print(f"Theoretical <phi|B_rho|phi> = {exp_B_rho_phi:.4f}")

# --- 3. Qiskit Simulation ---
exp_results = []
simulator = AerSimulator()
for i in range(2):
    psi_k_circuit = QuantumCircuit(1)
    if i == 1: psi_k_circuit.x(0)
    
    circuit = QuantumCircuit(1, 1)
    circuit.append(psi_k_circuit, [0])
    circuit.append(A, [0])
    circuit.append(phi_circuit.inverse(), [0])
    circuit.measure(0, 0)
    
    result = simulator.run(transpile(circuit, simulator), shots=10000).result()
    prob_0 = result.get_counts().get('0', 0) / 10000
    exp_results.append(prob_0)

simulated_rhs = pk[0] * exp_results[0] + pk[1] * exp_results[1]
print(f"Simulated average sum_k pk*|<phi|A|psi_k>|^2 = {simulated_rhs:.4f}")
\end{verbatim}

\end{document}